\documentclass[aps,prb,twocolumn,showpacs,groupedaddress,epsfig]{revtex4}
\usepackage{graphicx}
\usepackage{dcolumn}
\usepackage{amsmath}
\usepackage{longtable}
\usepackage{multirow}

\begin{document}

\title{Chemical engineering of adamantane by lithium functionalization: A first-principles density functional theory study}

\author{Ahmad Ranjbar$^1$} 
\email{ranjbar@imr.edu}
\author{Mohammad Khazaei$^1$}
\author{Natarajan Sathiyamoorthy Venkataramanan$^2$ }
\author{Hoonkyung Lee$^{3,4}$}
\author{Yoshiyuki Kawazoe$^1$}
\affiliation{$^1$Institute for Materials Research, Tohoku University, Sendai 980-8577, Japan}
\affiliation{$^2$California State University, College of Science, Hayward, CA 94542, USA}
\affiliation{$^3$Department of Physics, University of California, Berkeley, California 94720, USA}
\affiliation{$^4$Materials Sciences Division, Lawrence Berkeley National Laboratory, Berkeley, California 94720, USA}

\date{\today}

\begin{abstract}
Using first-principle density functional theory, we investigated the hydrogen storage capacity of Li functionalized adamantane. We showed that if one of the acidic hydrogen atoms of adamantane is replaced by Li/Li$^+$, the resulting complex is activated and ready to adsorb hydrogen molecules at a high gravimetric weight percent of around $\sim$7.0 \%. Due to polarization of hydrogen molecules under the induced electric field generated by positively charged Li/Li$^+$, they are adsorbed on ADM.Li/Li$^+$  complexes with an average binding energy of $\sim$-0.15 eV/H$_2$, desirable for hydrogen storage applications. We also examined the possibility of the replacement of a larger number of acidic hydrogen atoms of adamantane by Li/Li$^+$ and the possibility of aggregations of formed complexes in experiments. The stabilities of the proposed structures were investigated by calculating vibrational spectra and doing MD simulations. 
\end{abstract}


\maketitle

\section{Introduction}
Hydrogen storage is of great interest as environmentally clean and efficient fuels required for future energy applications. Several pioneering strategies have been developed and significant performances have been achieved for hydrogen storage, including chemisorption of dihydrogen in the form of light metal hydrides, metal nitrides and imides, and physisorption of dihydrogen onto carbon, clathrate hydrates, and porous network materials such as carbon nanotubes, zeolites, and metal-organic framework (MOF) materials~\cite{Schlapbach}. However, hydrogen storage in these systems requires either high pressure or very low temperature, or both, thus severely limiting the applicability for mobile applications. Furthermore, their synthesis at bulk scale in a discrete and tailored fashion in a high yield is quite difficult~\cite{Han, Thallapally}. Thus, the synthesis of functional materials with high hydrogen uptake and delivery under safe and ambient conditions remain a key challenge for establishing a hydrogen economy. 

To improve the capability of hydrogen storage in materials, it has been suggested that they should be doped with transition metal impurity or alkali metals. Generally, however, metal impurities bind to carbon surfaces weakly and they undergo aggregation after several subsequent fueling cycles~\cite{ramanan1,ramanan2}. Also synthesis of lithium doped materials leads to the agglomeration of Li atoms, resulting in uneven binding of Li on the metal surface, thereby reducing the storage capacity of the materials. To date, no material that consists of high Li content with ultra high hydrogen storage capacity has been reported~\cite{Blomqvist,vanden}. Hence, theoretical suggestions and speculations thus far proposed have not yielded experimental or technological methods for large-scale production.    
Diamondoids are hydrocarbons with cubic-diamond cage structures that offer valuable chemical building blocks for new materials~\cite{Heagy}. Since adamantane (ADM)  can be isolated in large amounts from crude oil, research activities focusing on such compounds as novel materials are rapidly emerging. Experimental and theoretical studies on ADM showed that these molecules undergo self-assembly on metal surfaces or into molecular crystals in vacuum with high porosity~\cite{Dahl,William-Clay,Simon,May, Piekarczyk,Zhang-PRB,Herman,xue}. Furthermore, a synthetic approach to functionalize the ADM molecule with Li in high purity and yield can be realized by the addition of bases such as organolithium~\cite{Molle} and the product has been well characterized using various spectroscopic methods~\cite{Scheler}.

 In this paper we have considered the hydrogen storage application of Li functionalized ADM. Our first-principles calculations show that by chemical substitutions of hydrogen atoms of ADM with Li/Li$^+$, it becomes activated and ready to adsorb hydrogen molecules with binding energy of -0.1 to -0.2 eV, with high gravimetric weight percents of more than 7.0 \%. The Department Of Energy (DOE) targets for economical hydrogen storage materials specified for 2010 are 6.0 wt \% and 45 kg/m$^3$ for gravimetric and volumetric capacities, respectively~\cite{DOE1,DOE2}. As another important criterion, recently, Bhatia and Myers studied the optimum thermodynamic conditions for hydrogen adsorption, employing the Langmuir equation and derived relationships between the operating pressure of a storage tank and the enthalpy of adsorption required for storage near room temperature. They have found that the average optimal adsorption enthalpy should be in the range of 0.1-0.2 eV/H$_2$~\cite{Jena,Bhatia}. It is seen that our designed materials satisfy the criteria for the gravimetric density of hydrogen in storage media by the DOE as well as above predicted optimal energy window for hydrogen adsorption. Therefore diamondoids-Li/Li$^+$ complexes might be good candidates for hydrogen storage materials.

\section{Computational method}
Density functional theory and \textit{ab-initio} molecular dynamics calculations were performed with the Vienna ab-initio Simulation Package (\textsc{VASP})~\cite{VASP} using the projector augmented wave (PAW) method~\cite{PAW} to describe the ion-electron interactions. Electron exchange-correlation functionals were represented with the generalized gradient approximation (GGA), with two nonlocal corrections, Perdew and Wang (PW91)~\cite{PW91}  and Perdew-Burke-Ernzerhof (PBE)~\cite{PBE} functionals. We used a simple large cubic cell with a lattice parameter of 35 {\AA}. For all calculations, a plane-wave expansion cut-off of 400 eV was used and the surface Brillouin zone integration was calculated using a $\Gamma$ point due to the use of large cubic cell size and cluster-type calculation. The structures were fully optimized until the magnitude of force on each ion becames less than 0.005 eV/{\AA} . The convergence criterion on the total energy  was set to 1$\times$10$^{-5}$ eV. Molecular dynamics calculations were carried out at a constant temperature for 48 ps with a time scale of 2 fs to determine the stability of the alkali functionalized ADM molecules. The calculations with hybrid meta GGA functional, (M05-2X)~\cite{M05-2X}, and the second-order M{\O}ller-Plesset perturbation theory (MP2)~\cite{MP2} were carried out as implemented in the \textsc{GAUSSIAN09} program package~\cite{gaussian09}. DFT and MP2 calculations  were carried out using the 6-31+G(d, p) and 6-311+G(2df, p) basis sets~\cite{MP2,Basis-set}. The choice of M05-2X/6-311+G(2df, p) for DFT calculations is justified as a compromise between reliable results and a reasonable computational cost compared to MP2 method. 

The consecutive binding energy of Li, the required energy to separate Li from ADM.Li$_m$ complex, is calculated subsequently as $E_b=E_{\textrm{ADM.Li}_{m+1}}-E_{\textrm{ADM.Li}_m}-E_{\textrm{Li}}$, where $E_{\textrm{ADM.Li}_{m+1}}$,  $E_{\textrm{ADM.Li}_m} $ $(m \geq 1)$ are the total energies of ADM with $m$+1 and $m$ substituted Li atoms, respectively. $E_{\textrm{Li}}$ is the total energy of a Li atom. The binding energy of Li in the ADM.Li is calculated as $E_b=E_{\textrm{ADM.Li}}-E_{\textrm{ADM}^*}-E_{\textrm{Li}}$, where $E_{\textrm{ADM.Li}}$ and $E_{\textrm{ADM}^*}$ are the total energy of optimized structures of ADM.Li and ADM with one detached acidic hydrogen atom, respectively. Similarly, the binding energy of the Li$^+$ in ADM.Li$^+$ is calculated as $E_b=E_{\textrm{ADM.Li}^+}-E_{\textrm{ADM}^*}-E_{\textrm{Li}^+}$. 

The binding energy per hydrogen molecule, $E_b$, was calculated using 
\begin{equation}
\label{Eb}
E_b= \frac{1}{n}\left[E_{\textrm{ADM.Li}_m-(\textrm{H}_2)_n} - E_{\textrm{ADM.Li}_m} - nE_{\textrm{H}_2}\right],
\end{equation}
 where $E_{\textrm{ADM.Li}_m-(\textrm{H}_2)_n}$, $E_{\textrm{ADM.Li}_m}$ and  $E_{\textrm{H}_2}$ are the total energies obtained for alkali atom-doped ADM containing $n$ hydrogen molecule(s) and $m$ lithium atom(s), the alkali atom-doped ADM system and  an isolated H$_2$ molecule being located in the same supercell.

 In our study, the excess and depletion charge or difference charge density $\triangle\rho$ is estimated by
$\triangle\rho=\rho(\textrm{ADM.Li})-\rho(\textrm{ADM}^*)-\rho(\textrm{Li})$, where $\rho(\textrm{ADM.Li})$  stands for the charge density of full relaxed structure of ADM.Li. $\rho(\textrm{ADM}^*)$ and $\rho(\textrm{Li})$ are the charge densities of ADM$^*$ (ADM.Li with detached Li) and Li atom, which are obtained from two separate single point energy calculations without any relaxation while the positions of all the atoms in ADM$^*$ and Li are kept fixed as the positions of their corresponding atoms in ADM.Li.

 \section{Results and discussion}

Before evaluating the hydrogen storage property of chemically modified diamondoid-based complexes, as an example of pristine diamondoid structures, we examined the hydrogen storage property of ADM using various density functional approaches. In all our calculations, it was observed that due to the very high stability of ADM~\cite {saani}, the binding energy of hydrogen molecules on such compounds was very small, i.e., on the order of $\sim-0.001$ eV, which is out of the energy range, from -0.1 to -0.2 eV, desirable for reversible H$_2$ adsorption/desorption near room temperature for hydrogen storage applications~\cite{Jena,Bhatia}. Hence, pristine diamondoids are not considered to be good candidates for hydrogen storage applications. However, experimentally it has been shown that by treating the surfaces of diamondoids with chemical solutions~\cite{fokin,maison,krueger,schwertfeger, schreiner}, the hydrogen atoms of diamondoids can be selectively replaced by other compounds or elements, resulting in significant changes in their electronic structures~\cite{fokin2,landt}. Among the experimentally formed diamondoid-based complexes, one modified with Li/Li$^+$~\cite{Molle,kraus,adcock} can attract attention for hydrogen storage applications and it would be worthwhile to consider their storage properties in detail. It should be noted that it has already been shown theoretically that Li can improve the hydrogen storage property of carbon materials~\cite{sun,deng,cabria,liu2,khazaei,froudakis}. However, experimentally it is still controversial. The wide range of experimental results for hydrogen storage improvements after Li doping, from less than 1 wt\% to several tens of weight percentages at moderate pressures and temperatures ~\cite{yang,chen,pinkerton}, are  mainly attributed to sample preparation issues~\cite{chen,yang,pinkerton}. 

\begin{figure}[pt]
 \begin{center}
   \includegraphics[width=3.4 in]{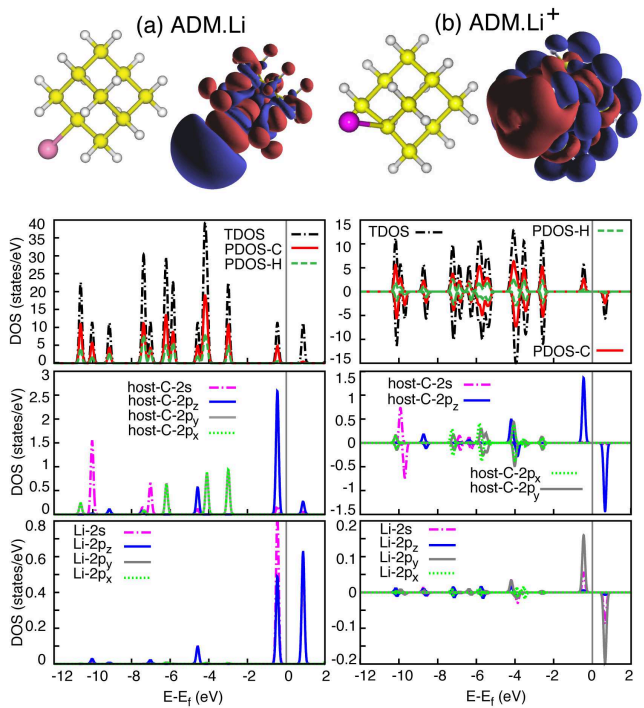}
  \end{center}
  \caption{(Color online)  DFT-optimized structures of (a) ADM.Li, (b) ADM.Li$^+$ and their excess (red)$-$depletion (blue) charges and the related total and projected densities of states.}
  \label{fig:ex-dep}
\end{figure}

Since ADM has four acidic hydrogen atoms, it can possibly be modified with four Li atoms/ions. Our calculations show that in ADM molecules, more than one acidic hydrogen cannot be replaced by a Li$^+$ because the structure of ADM is significantly deformed such that the binding energy of the second substituted Li$^+$ becomes positive, +0.15 eV . Figure~\ref {fig:ex-dep} shows the most stable configurations of ADM when one of its acidic hydrogen atoms is replaced by Li/Li$^+$. From  Fig.~\ref {fig:ex-dep}, it is observed that in ADM.Li, Li forms a bond with the host carbon atom, similar to other acidic hydrogen atoms, while in ADM.Li$^+$, the Li$^+$  is tilted toward the neighboring hydrogen atoms. There is another possible configuration for ADM.Li (ADM.Li$^+$) with a tilted (straight) Li$-$C (Li$^+-$C) bond, which is $\sim$0.05 eV less stable than the one shown in Fig.~\ref {fig:ex-dep}. Therefore, it is expected that both configurations with straight and tilted Li/Li$^+-$C bonds can be formed in ambient experimental conditions. 
Table~\ref {tab:Li-binding} shows the consecutive binding energy of Li/Li$^+$  in ADM.Li$_m$ (m=1-4) and ADM.Li$^+$. The trend of the changes of binding energies of Li to ADM.Li$_m$ are observed to be the same using different methods, PW91, PBE, M05-2X, and MP2; by increasing the number of Li atoms, the binding energy of Li to ADM.Li$_m$ decreases. Furthermore, our calculations show that the binding energy of Li in ADM.Li is almost same as the binding energy of Li$^+$  in the ADM.Li$^+$ complex. To confirm the stability of ADM.Li$_m$/Li$^+$, we have calculated the vibrational spectra of our designed structures. The obtained results show the absence of any imaginary frequency, indicating that the optimized structures are real minima. We also performed MD simulations at 400 K for 48 ps and observed no detachment of Li/Li$^+$  from the structures and no Li aggregation on an individual ADM.Li$_m$ complex. As previously mentioned, Li/Li$^+$ has two nearly isomer configurations with straight and tilted Li/Li$^+-$C bonds; as a result, during the MD simulations at high temperature, Li/Li$^+$ has pendulum movement between two local minima. The possibility of formation of ADM.Na and ADM.K complexes was also considered, and the binding energy of Li to ADM (see Table~\ref {tab:Li-binding}) was found to be larger than that of Na (-0.86 eV) or K (-0.76 eV) atoms to ADM. Hence, we focused our examination on just ADM.Li$_m$/Li$^+$ complexes.

\begin{table}[pt]
\caption{\label{tab:Li-binding}$E_{b}$, the calculated consecutive binding energies of Li /Li$^+$ in ADM.Li$_m$/Li$^+$ complexes. Single point MP2 calculations were done for the optimized structures obtained from M05-2X/6-311+G(2df,p).}
\begin{ruledtabular}
\begin{tabular}{llllcccccccc}
 &\multicolumn{2}{c}{PW91\footnotemark[1]}&PBE\footnotemark[1]&\multicolumn{2}{c}{M05-2X\footnotemark[2]}&M05-2X\footnotemark[3]&MP2\footnotemark[4]\\
 \cline{2-3}\cline{4-4}\cline{5-6}\cline{7-7}\cline{8-8}
 \\
  Cluster&$E_{b}$&$\bar{d}_{Li-C}$&$E_{b}$ \footnotemark[1]&$E_{b}$&$\bar{d}_{Li-C}$&$E_{b}$&$E_{b}$\\
\hline
 ADM.Li&-1.46&2.028&-1.29&-1.57&2.010&-1.51&-1.59\\
 ADM.Li$_2$&-1.16&-2.050&-1.02&-1.36&2.023&-1.30&-1.36\\
 ADM.Li$_3$&-0.94&2.075&-0.84&-1.17&2.041&-1.14&-1.24\\
 ADM.Li$_4$&-0.71&2.094&-0.62&-1.06&2.059&-1.05&-1.25\\
ADM.Li$^+$&-1.62&2.082&-1.60&-1.52&2.038&-1.46&-1.43\
 \footnotetext[1]{Basis set: plan-wave.}
 \footnotetext[2]{DFT cluster method, basis set: 6-311+G(2df, p).}
 \footnotetext[3]{DFT cluster method, basis set: 6-31+G(d, p).}
 \footnotetext[2]{Basis set: 6-311+G(2df, p).}

 \end{tabular}
 \end{ruledtabular}
 \end{table}

To consider the nature of Li/Li$^+$  bonding in ADM.Li/Li$^+$, we performed the excess and depletion charge analysis and plotted their total and projected density of states, as shown in Fig.~\ref{fig:ex-dep}. The projected density of states was plotted for the Li/Li$^+$ and the host carbon atoms. From excess and depletion charge analysis, in ADM.Li, Li was found to become positively charged by donating its $2s$ electron to the cluster, particularly the host carbon atom and all the hydrogen atoms in the complex. In ADM.Li$^+$, Li$^+$  becomes less positively charged by accepting electrons, mainly from the hydrogen atoms of the cluster. From the projected density of state calculations, the electrons donated to Li$^+$ were observed to move not only to its $2s$ orbital, but also to its $2p_y$ orbital. In the ADM.Li complex, the $2s$ orbital of Li is partially hybridized with the $2p_z$ orbital of the host carbon atom (see PDOS peaks at -0.25 eV) and makes a $sp^3$-like bond with it. In ADM.Li$^+$, the $2s$ and $2p_y$ orbitals of Li$^+$ are partially hybridized with the $2p_z$ orbital of the host carbon atom (see the states close to Fermi energy). From our calculations, it is concluded that the bonding nature of Li/Li$^+$ in ADM.Li/Li$^+$ is predominantly ionic and partially covalent.      

\begin{figure}[pt]
 \begin{center}
   \includegraphics[width=3.2 in]{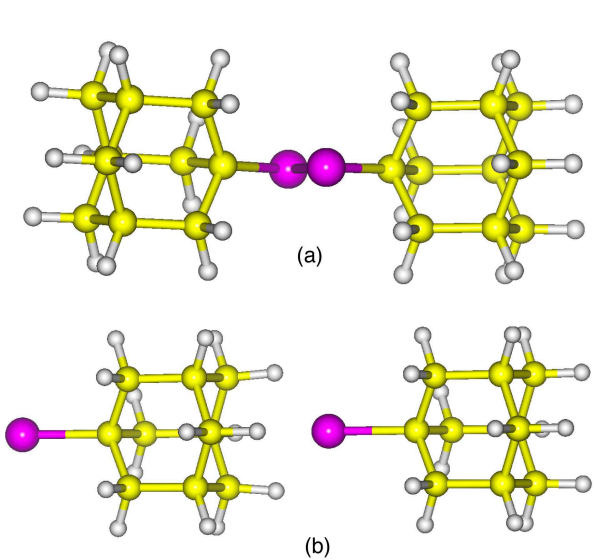}
  \end{center}
  \caption{(Color online)  DFT-optimized structures of (a) dimer and (b) chain-like configurations of two attached ADM.Li complexes.}
  \label{fig:dimer}
\end{figure}

Before considering the hydrogen storage properties of ADM.Li/Li$^+$  complexes, let us consider the possibility of clustering of these new structures when two or more of them get close to each other. The ADM.Li complexes can form dimers without any energy barrier when they get close to each other along their heads with Li. From our projected density of states analysis, when two ADM.Li approach head-to-head (see Fig.~\ref{fig:dimer}(a)), it was observed that the $2p_y$ orbitals of two Li atoms are hybridized with a binding energy of -1.95 eV. Our calculations show that ADM.Li complexes can also become connected as a chain-like structure with a binding energy of -0.33 eV, see Fig.~\ref{fig:dimer}(b). When the Li at the head of one of the ADM.Li gets close to the ending hydrogen atoms of the other complex, they are bonded due to their electrostatic interaction between the positively charged Li of the first complex and the negatively charged hydrogen atoms of the second complex; see the excess-depletion charges in Fig.~\ref{fig:ex-dep}. For the chain formation, there is also no energy barrier. However, our energy calculations show that the dimer configuration is much more stable than the chain structure. Therefore, it is expected that dimers would be more abundant than the chains in the experiment. 
In the case of ADM.Li$^+$, no clustering was observed. This is because in this type of complex, either Li or hydrogen atoms are positively charged. Hence, there is a strong repulsion between two ADM.Li$^+$  complexes that keeps them away from each other and prevents their clustering. It is worth mentioning that Xue and Mansoori have recently found that ADM.Na$^+$ complexes are self-assembled like a molecular crystal~\cite{xue} by performing MD simulations on 125 ADM.Na$^+$ complexes. The vacant spaces between the complexes may make it possible to store hydrogen molecules in high gravimetric weight percentages. Moreover, recently, experimentalists have been able to prepare the positively charged alkali metal doped MOF systems via electrochemical reduction~\cite{Li-reduction1,Li-reduction2}. They proved that the hydrogen storage capacity increases after introducing the alkali metal charge cations. Our studied system is an organic molecule, which can be easily charged electrochemically than MOFÕs. Hence we believe that ADM.Li$^+$  is a superior to ADM.Li as a candidate for hydrogen storage. However, we would like to continue evaluation of both ADM.Li and Li$^+$  structures, because recent advances in modifications of ADM in experiments may make it possible to separate ADM.Li$_m$ nano-particles from each other. 

\begin{figure}[pt]
 \begin{center}
   \includegraphics[width=3.3 in]{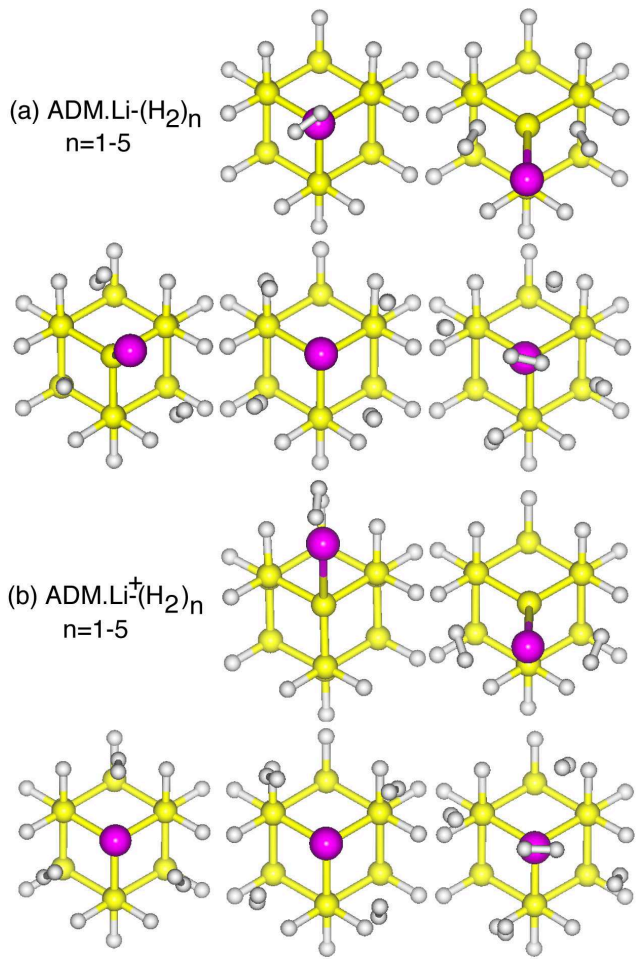}
  \end{center}
  \caption{(Color online)  DFT-optimized structures of (a) ADM.Li and (b) ADM.Li$^+$ when one to five hydrogen molecules are adsorbed.}
  \label{fig:storage}
\end{figure}

The next step of our study was to consider the hydrogen storage property of ADM.Li$^+$  and ADM.Li$_m$ complexes. As summarized in Table~\ref{tab:H$_2$-binding} and as seen from Fig.~\ref{fig:storage}, each Li/Li$^+$ site adsorbs a maximum of five H$_2$ molecules. Therefore, it can be predicted that the gravimetric weight percentage of hydrogen storage for ADM.Li$^+$  is $\sim$7.0 \% and  between 7.0-20.0 \% for ADM.Li$_m$ complexes, if experimentalists can find a way to prevent their clustering. If clustering occurs the storage properties of ADM.Li$_m$ will be less than the above mentioned value. 

\begin{table*}[pt]
\caption{\label{tab:H$_2$-binding}$E_{b}$ (in eV), $\bar{L}_{H-H}$, $\bar{d}_{H_2-Li}$, and $\bar{d}_{Li-C}$ (in \AA) are the calculated binding energies of adsorbed hydrogen molecules, bond length average of hydrogen molecule(s), average distance between the center of hydrogen molecule(s) and Li/Li$^+$, and average bond distance of Li and host carbon atoms in different methods, respectively.}
\begin{ruledtabular}
\begin{tabular}{llllllllcccccccccccc}
 &\multicolumn{4}{c}{PW91\footnotemark[1]}&PBE\footnotemark[1]&\multicolumn{4}{c}{M05-2X\footnotemark[2]}&M05-2X\footnotemark[3]&MP2\footnotemark[4]\\
 \cline{2-5}\cline{6-6}\cline{7-10}\cline{11-11}\cline{12-12}
 \\
  Cluster&$E_{b}$&$\bar{L}_{H-H}$&$\bar{d}_{H_2-Li}$&$\bar{d}_{Li-C}$&$E_{b}$&$E_{b}$&$\bar{L}_{H-H}$&$\bar{d}_{H_2-Li}$&$\bar{d}_{Li-C}$&$E_{b}$&$E_{b}$\\
\hline
 \\
  ADM.Li-(H$_2)_1$&-0.10&0.753&2.184&2.020&-0.10&-0.11&0.743&2.149&2.008&-0.11&-0.11\\
  ADM.Li-(H$_2)_2$&-0.20&0.788&1.788&2.007&-0.20&-0.15&0.760&1.868&1.948&-0.12&-0.10\\
  ADM.Li-(H$_2)_3$&-0.15&0.770&1.882&2.049&-0.14&-0.15&0.749&1.965&2.011&-0.14&-0.12\\
  ADM.Li-(H$_2)_4$&-0.14&0.763&1.966&2.088&-0.13&-0.15&0.748&2.034&2.035&-0.14&-0.12\\
  ADM.Li-(H$_2)_5$&-0.11&0.760&2.285&2.085&-0.11&-0.13&0.746&2.197&2.023&-0.13&-0.10\\
  ADM.Li$_2$-(H$_2)_8$&-0.14&0.766&1.943&2.104&-0.13&-0.14&0.750&1.998&2.042&-0.14&-0.11\\
  ADM.Li$_3$-(H$_2)_{12}$&-0.14&0.767&1.922&2.119&-0.13&-0.14&0.751&1.965&2.052&-0.14&-0.11\\
  ADM.Li$_4$-(H$_2)_{16}$&-0.14&0.768&1.908&2.134&-0.13&-0.14&0.753&1.937&2.065&-0.13&-0.10\\
  ADM.Li$_4$-(H$_2)_{20}$&-0.12&0.765&2.333&2.135&-0.11&-0.12&0.750&2.221&2.048&-0.11&-0.08\\
  ADM.Li$^+$-(H$_2)_1$&-0.23&0.757&2.022&2.090&-0.21&-0.21&0.746&2.042&2.046&-0.17&-0.21\\
  ADM.Li$^+$-(H$_2)_2$&-0.21&0.757&2.025&2.146&-0.19&-0.20&0.746&2.028&2.113&-0.16&-0.20\\
  ADM.Li$^+$-(H$_2)_3$&-0.19&0.756&2.049&2.184&-0.17&-0.19&0.746&2.031&2.145&-0.15&-0.19\\
  ADM.Li$^+$-(H$_2)_4$&-0.17&0.755&2.139&2.223&-0.15&-0.17&0.744&2.124&2.187&-0.14&-0.17\\
  ADM.Li$^+$-(H$_2)_5$&-0.15&0.755&2.236&2.270&-0.13&-0.16&0.744&2.192&2.241&-0.13&-0.15\\
  
  H$_2$&&0.749&&&&&0.739&&&&\\
 
  \footnotetext[1]{Basis set: plan-wave}
  \footnotetext[2]{DFT cluster method, basis set: 6-311+G(2df, p).}
 \footnotetext[3]{DFT cluster method, basis set: 6-31+G(d, p).}
 \footnotetext[4]{Basis set: 6-311+G(2df, p).}

 \end{tabular}
 \end{ruledtabular}
 \end{table*}

Now let us focus on the hydrogen adsorption on  Li/Li$^+$ functionalized ADM. The calculated binding energies along with the bond parameters for the hydrogen molecules using various functionals and basis sets are given in Table~\ref{tab:H$_2$-binding}. From the table, it is seen that in our case study, both pure and hybrid functionals provide binding energies close to the more accurate MP2 method. It is observed that the binding energies of hydrogen molecules on ADM.Li$_m$/Li$^+$  are on the order of -0.1 to -0.23 eV, which is very good for hydrogen storage applications. Furthermore, calculated binding energies for the cationic Li$^+$ are higher than those of the neutral Li-doped ADM system. The calculated Li-H distance increases with an increase in the number of hydrogen molecules. The small changes in binding energy of hydrogen molecules can be attributed to various reasons: the amount of positive charges on Li~\cite{ataca1}, the distances between H$_2$ molecules and Li, the Li$-$C bond distance~\cite{huang}, the interaction between the hydrogen molecules~\cite{ao,ataca2}, the strength of hybridization of hydrogen molecules with Li, etc. It is observed that the adsorbed H$_2$ molecules are located at distances of $\sim$2.1 {\AA} of Li/Li$^+$. It is also seen that the first H$_2$ is adsorbed on top of the Li/Li$^+$ (see Fig.~\ref{fig:storage}), but when the second, third, or fourth H$_2$ molecules are adsorbed, they prefer to move to the lateral side of Li/Li$^+$. As observed from Fig.~\ref{fig:storage}, when the hydrogen molecules are adsorbed on the ADM.Li/Li$^+$ complexes, the position of Li/Li$^+$ changes between their two local minima with a straight or tilted Li/Li$^+-$C bond in order to reduce the steric repulsion between the adsorbed hydrogen molecules and the hydrogen atoms of the ADM.Li/Li$^+$ structures.

\begin{figure}[pt]
 \begin{center}
   \includegraphics[width=3.5 in]{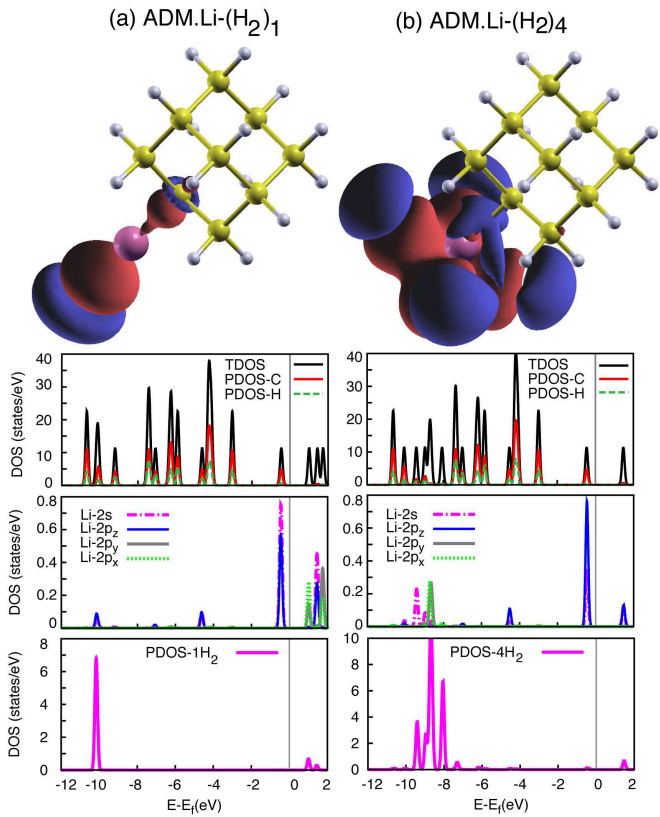}
  \end{center}
  \caption{(Color online)  Excess (red)- depletion (blue) charge iso-surfaces and total and projected densities of states for  (a) ADM.Li-(H$_2)_1$ and (b) ADM.Li-(H$_2)_4$.}
  \label{fig:dos}
\end{figure}

\begin{figure}[pt]
 \begin{center}
   \includegraphics[width=3.5 in]{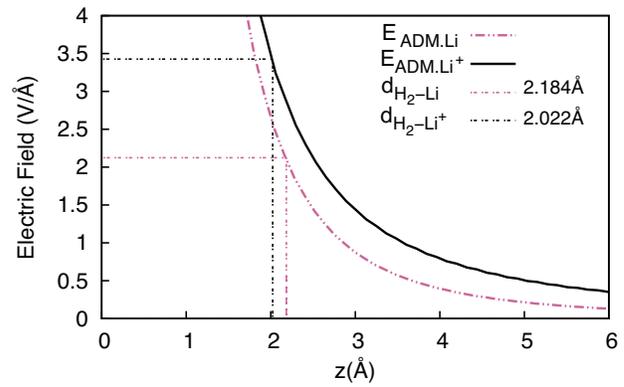}
  \end{center}
  \caption{(Color online) Induced electric field in the direction of the Li/Li$^+-$C bond of ADM.Li/Li$^+$. Li/Li$^+$ located at the origin. The dotted lines indicate
  the center of the adsorbed hydrogen molecule.}
  \label{fig:Electric}
\end{figure}

To find the nature of bonding between H$_2$ and Li/Li$^+$, as an example, we considered the excess and depletion charge and the projected density of states for ADM.Li when one or four H$_2$ molecules are adsorbed on a Li atom (see Fig.~\ref{fig:dos}). From the excess and depletion charge analysis, positively charged Li was shown to polarize H$_2$ molecules under its induced electric field. Due to this polarization, there is a small bond elongation for the H$_2$ molecules, as seen from Table~\ref{tab:H$_2$-binding}, but no dissociation of H$_2$ molecules. Density of states analysis indicated that when the number of H$_2$ molecules increases from one to four, they start to interact with each other such that the states related to H$_2$ molecules are broadened~\cite{ataca2}, between $\sim$-8.0 eV and -10.0 eV. Regarding the ADM.Li-(H$_2)_1$, as shown in Fig.~\ref{fig:dos}, the $1s$ orbital of H$_2$ molecule is slightly hybridized with the $p_z$ orbital of the Li atom at -10 eV. When the number of H$_2$ molecules changes to four, the $1s$ orbitals of H$_2$ molecules prefer to interact with the $p_x$ and $p_y$ orbitals of Li (at energies between $\sim$-8.0 eV and -8.5 eV) instead of its $p_z$ orbital. Hence, when the number of H$_2$ molecules increases, they prefer to locate in the lateral positions of Li rather than in the top position. To better understand how the electric filed induced by positively charged Li affects the binding energies of H$_2$ molecules, we plotted the magnitude of the induced electric field along the Li/Li$^+-$C bond for ADM.Li/Li$^+$, shown in Fig~.\ref{fig:Electric}. As shown by the figure, the amount of generated electric field at the center of adsorbed H$_2$ molecule on Li/Li$^+$ is on the order of 2.1/3.4 (V/\AA). The polarizability of a hydrogen molecule along ($\alpha_\|$) and perpendicular ($\alpha_\bot$) to the hydrogen molecule axis in an external electric field are 6.3 a.u. and 4.85 a.u., respectively~\cite{saika,mura,Jackson}. The adsorption energy of hydrogen in such electric fields is estimated to be $E_b$$\sim$-1/2 $\vec{P}.\vec{E}$$_{ext}$$\sim$-1/2$\alpha$ $E_{ext}^2$$\sim$-0.11/-0.29 eV. These values are very close to ones reported in Table~\ref{tab:H$_2$-binding}. Therefore, it is expected that the electrostatic interactions between H$_2$ molecules and Li make larger contributions to the binding energy of H$_2$ molecules on ADM.Li/Li$^+$  complexes than their hybridizations. It is worth mentioning, as seen from Table~\ref{tab:H$_2$-binding}, that the average of adsorption energy of H$_2$ molecules on ADM.Li$_m$ does not change significantly when the number of Li sites increases. This indicates that in such complexes, the interaction between hydrogen molecules and positively charged Li is still electrostatic. It should be mentioned that Yoon \textit{et al.}~\cite{yoon} have used a similar analysis and successfully explained why hydrogen molecules can be adsorbed on charged and doped carbon fullerenes. Also, Zhou \textit{et al.}~\cite{zhou} have recently shown that under an applied electric field, hydrogen storage property of a boron nitride sheet is substantially improved due to polarization of the hydrogen molecules as well as the substrate.

\begin{figure}[pt]
 \begin{center}
   \includegraphics[width=3.3 in]{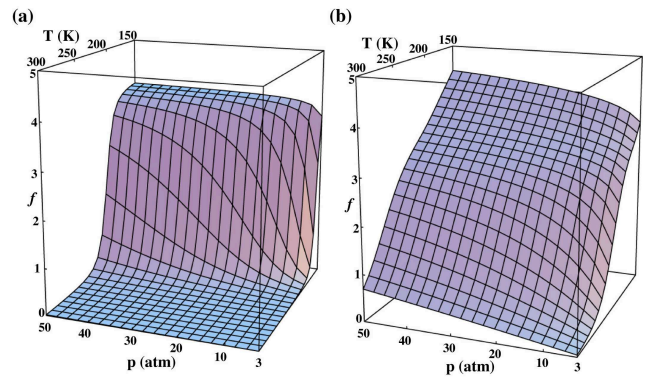}
  \end{center}
  \caption{(Color online) Occupation number as a function of the pressure and temperature on (a) ADM.Li and (b) ADM.Li$^+$.}
  \label{fig:f-function}
\end{figure}

To investigate the thermodynamics of adsorption of H$_2$ molecules on ADM.Li and ADM.Li$^+$, the occupation number of H$_2$ molecules per site (Li/Li$^+$ atom) was calculated as a function of the pressure and temperature using the following formula~\cite{Lee,Lee2}:  
\begin{equation}
\label{f-function }
f=\frac{\displaystyle\sum_{n=0} g_{n}ne^{n(\mu-\epsilon_{n})/kT}}{\displaystyle\sum_{n=0} g_{n}e^{n(\mu-\epsilon_{n})/kT}}
\end{equation}
where $\mu$ is the chemical potential of the H$_2$ gas, $\epsilon_n(<0)$ and $g_n$ are the (average) binding energy of the H$_2$ molecules and the degeneracy of the configuration for a given adsorption number of the H$_2$ molecules $n$, respectively and, $k$ and $T$ are the Boltzmann constant and the temperature, respectively. Figure~\ref{fig:f-function}  shows the occupation number of H$_2$ molecules on the Li and Li$^+$ atoms as a function of the pressure and temperature where the experimental chemical potential of H$_2$ gas~\cite{handbook} and the calculated binding energy ($\epsilon_n$) obtained from MP2 calculations were used. The occupation number of H$_2$ molecules $f$ at 150 K and 30 atm in both cases is $\sim$4, as shown in Figure~\ref{fig:f-function}. This is attributed to the Gibbs factor ($e^{4(\mu-\epsilon_4)/kT}$) for the binding of four H$_2$ molecule, which dominates at 150 K and 30 atm ($\mu=-0.08$ eV, $\epsilon_4$ on the Li and Li$^+$ is -0.12 and -0.17 eV, respectively). The number goes to zero at room temperature ($\mu=\sim-0.32$ eV). Therefore, this analysis shows that the ADM.Li/Li$^+$ structures may have considerable potential as high-capacity hydrogen storage materials.

\section{Conclusions} 
Using different first-principles approaches, we have shown the possibility of formation of ADM.Li and ADM.Li$^+$. It is predicted that ADM.Li$^+$ structures will not be clustered under ambient conditions, while there is a tendency for clustering of ADM.Li complexes. We found that each Li$^+$ is capable of holding five hydrogen molecules in molecular form. Binding energy of the hydrogen molecules was in the range of -0.15 eV to -0.23 eV, which was due to the induced electric field created by the Li$^+$ ion.  Furthermore, the calculated occupancy number reveals that ADM.Li/Li$^+$ to be ideal candidates for hydrogen storage. Thus, ADM.Li$^+$ is expected to be a superior candidate for use as a hydrogen storage medium, as hydrogen molecules can be adsorbed or desorbed from this complex at pressures and temperatures accessible in recent technologies.

\begin{acknowledgments}
The authors gratefully acknowledge the Center for Computational Materials Science at the Institute for Materials Research for use of the Hitachi SR11000 (Model K2) supercomputer system. M. Khazaei thanks the Japan Society for the Promotion of Science (JSPS) for financial support.
\end{acknowledgments}

\bibliography{basename of .bib file}

\end{document}